\documentclass[11pt, amsmath, a4paper]{article}

\newcommand {\basel}[2]{#1_{_{#2}}}

\def \lspace  {\hspace*{-0.1cm}}

\def \tab  {\hspace*{0.5cm}}
\def \ltab  {\hspace*{-0.5cm}}

\def \shiftright  {\hspace*{2.0cm}}
\def \shiftleft  {\hspace*{-2.0cm}}

\newcommand   \qed   {\hfill{}$\Box$}

\usepackage{amsfonts}
\usepackage{amssymb}
\usepackage{mathrsfs}
\usepackage{graphicx}

\title{Fast Parallel Integer Adder in Binary Representation}

\author{\bf{Duggirala Meher Krishna}\\
{\small{Gayatri Vidya Parishad College of Engineering (Autonomous)}} \\
{\small{Madhurawada, VISAKHAPATNAM -- 530 048, Andhra Pradesh, India}} \\
 {\small{E-mail ~: \tab duggiralameherkrishna@gmail.com}}\\
 \\
 and \\
 \\
\bf{Duggirala Ravi}\\
{\small{Gayatri Vidya Parishad College of Engineering (Autonomous)}} \\
{\small{Madhurawada, VISAKHAPATNAM -- 530 048, Andhra Pradesh, India}} \\
\shiftleft \tab {\small{E-mail ~: \tab ravi@gvpce.ac.in; \tab duggirala.ravi@yahoo.com};} \\
{\small{\shiftright  duggirala.ravi@rediffmail.com; \tab drdravi2000@yahoo.com}} 
}

\date{}

\begin{document}

\maketitle

\begin{abstract}
An integer adder for integers in the binary representation is one of the basic operations of any digital processor. For adding two integers of N  bits each, the serial adder takes as many clock ticks. For achieving higher speeds, parallel circuits are discussed in the literature, and these circuits usually operate in two levels. At the lower level, integers represented by blocks of smaller number of bits are added, and in a cascade of stages in the next level, the carries produced in previous addition operations are summed to the augends. These circuits perform addition of integers of $N$ bits in about ${\mathcal{O}}(\basel{\log}{2}N)$ number of clock ticks and ${\mathcal{O}}(N*\basel{\log}{2}N)$ space. In this paper, we describe a fast method and an improvement of it. The first attempt resembles the operation method of the merge sort algorithm, from which some important properties of carries produced in each stage are analysed and assimilated, resulting in a parallel adder that runs in about ${\mathcal{O}}(\basel{\log}{2}N)$ number of clock ticks and  ${\mathcal{O}}(N*\basel{\log}{2}N)$ space.  Then, the crucial insights are brought to fruition in an improved design, which takes 2 clock ticks to perform the addition operation, requiring only ${\mathcal{O}}(N^{2})$ space. The number of bits $N$ is chosen usually to be a positive integer power of $2$. The speedup is achieved by special purpose circuits for increment operations by $2^{i}$, for $0 \leq i \leq N-1$, each operation taking only a single clock tick to complete. The usefulness of this adder for multiplication operation is discussed. The standard multiplication method utilizes quantizer and $3$-bit to $2$-bit consolidation circuits to produce an integer that represents in binary the number of $1$s in a column corresponding to a place (weighted coefficient) of nonnegative integer power of $2$. The last two consolidated integers are added by an adder in the end. 
\end{abstract}

\section{Introduction}

Addition operation of integers represented in binary is a basic operation on most, if not all, modern digital processors. The sequential or serial circuit for performing addition of two $N$ bit integers takes $N$ clock ticks. For parallelization of the addition operation, the main issue is to find an efficient method to deal with the carry produced by addition operation of smaller number of bits. For various methods discussed in the literature, {\em{viz}}, Ripple carry adder or Carry propagate adder, Carry look-ahead adder,  Carry skip adder,  Manchester chain adder,  Carry select adders,  Prefix adders,  Multi-operand adder,  Carry save adder, Pipelined parallel adder, {\em{etc.}}, see [\cite{BR:1982} -- \cite{GHM:1987}]. These circuits can perform addition of integers of $N$ bits in about ${\mathcal{O}}(\basel{\log}{2}N)$ number of clock ticks and ${\mathcal{O}}(N*\basel{\log}{2}N)$  space (see \cite{ASCS:2015, KS:2015}).

In the next section, we present a $k$-stage cascade circuit, where $N = 2^{k}$, performing addition operation in only $k$ clock ticks, requiring $k*2^{(k-1)}-1$ space for the special purpose circuits for carry addition. The motivation of this work is to present a unified and simplified circuit that can achieve the same task as discussed in the literature. Moreover, some important insights are gained in the design of this circuit, in a first attempt, which are exploited for realizing an improved circuit that adds in constant time, {\em {i. e.}}, in $2$ time delays, but requiring only at most $\frac{N(N+1)}{2}$ space. Further improvements, including the application of the adder for fast multiplication of two integers represented in binary, are discussed towards the end of the article. The standard multiplication method utilizes quantizer and $3$-bit to $2$-bit consolidation circuits to produce an integer that represents in binary the number of $1$s in a column corresponding to a place (weighted coefficient) of nonnegative integer power of $2$. The last two consolidated integers are added by an adder in the end.

\section{Parallel Binary Adder}

The steps involved in a parallel adder, resembling the merge sort algorithm, are described in the following algorithm:

\paragraph{\underline{First Attempt Parallel Adder Circuit}}
\begin{enumerate}
\item	Let the number of bits in the integers be $N = 2^{k}$, for some positive integer $k$.
\item	Let  $\basel{a}{(N-1)} \basel{a}{(N-2)} \ldots \basel{a}{0}$ and $\basel{b}{(N-1)} \basel{b}{(N-2)} \ldots \basel{b}{0}$ be the input integers in the binary form, with the convention that the most significant bit is the leftmost (and the least significant bit the rightmost).
\item	Initially, compute $2^{(k-1)}$ sums of two bits each, $\basel{s}{1,\,  2*i+1}  \basel{s}{1, \,  2*i}$, and the corresponding carries $\basel{c}{1, \,  i}$, such that, the binary sequences $\basel{s}{1,\,  2*i+1}  \basel{s}{1, \,  2*i}$ are the two lesser significant bits obtained by adding $\basel{a}{2*i+1}  \basel{a}{2*i}$ and $\basel{b}{2*i+1}  \basel{b}{2*i}$ , with a carry bit  $\basel{c}{1,\, i}$, for $0 \leq i \leq 2^{(k-1)}-1 = \frac{N}{2}-1$; this operation is performed separately by $2^{(k-1)}$ many programmable logic arrays or associative memory units, which compute in parallel for each index $i$, where  $0 \leq i \leq 2^{(k-1)}-1$.

\item	For $l = 1, \, 2, \, \ldots, \, k-1$, in steps of $1$, in the ascending order, after $2^{(k-l)}$ the sums of  $2^{l}$ bits each, $\basel{s}{l, \,  i*2^{l}+2^{l}-1}  \basel{s}{l,  \, i*2^{l}+2^{l}-2} \, \ldots \, \basel{s}{l, \,  i*2^{l}}$, together with the carries $\basel{c}{l, \, i}$  , for $ 0 \leq i \leq 2^{(k-l)}-1$, the following increment operation is performed : the integer represented by the binary sequence $\basel{c}{l, \,  2*i+1}  \basel{s}{l, \,  (2*i+1)*2^{l}+2^{l}-1} \basel{s}{l, \,  (2*i+1)*2^{l}+2^{l}-2} \,  \ldots \, \basel{s}{l, \,  (2*i+1)*2^{l}}$  is incremented by  $\basel{c}{l, \,  2*i}$, to get the carry $\basel{c}{l+1,  \, i}$  and left half string of the sum  $\basel{s}{l+1, \,  i*2^{(l+1)}+2^{(l+1)}-1} \basel{s}{l+1, \,  i*2^{(l+1)}+2^{(l+1)}-2} \, \ldots \, \basel{s}{l+1, \,  i*2^{(l+1)}+2^{l}}$ ,  and the right half string of the sum  $\basel{s}{l+1, \,  i*2^{(l+1)}+2^{l}-1} \basel{s}{l+1, \,  i*2^{(l+1)}+2^{l}-2} \, \ldots \, \basel{s}{l+1, \,  i*2^{(l+1)}}$  is taken to be the bit string $\basel{s}{l,\,   (2*i)*2^{l}+2^{l}-1}  \basel{s}{l, \,  (2*i)*2^{l}+2^{l}-2} \, \ldots \, \basel{s}{l, \,  (2*i)*2^{l}}$; the left and right half bit strings are concatenated to get the binary string  $\basel{s}{l+1, \,  i*2^{(l+1)}+2^{(l+1)}-1} \basel{s}{l+1, \,  i*2^{(l+1)}+2^{(l+1)}-2} \, \ldots \, \basel{s}{l+1, \,  i*2^{(l+1)}}$, representing the sum, with the corresponding  carry $\basel{c}{l+1,\, i}$ as just obtained; this increment operation can be performed in a single clock tick by a special purpose circuit, which identifies the least index $j$, where $0 \leq j \leq 2^{l}$, such that all the least significant bits up to (but not including) index $j$ are $1$ and the bit with index $j$ is $0$, by means of $(2^{l}+1)$ AND-gates, implemented by negated NOR-gates, and instantly complements the bits with index $j$ upto the least significant bit; if $\basel{c}{l, \,  2*i+1}$ is $0$, then there is one such index $j$, and if $\basel{c}{l, \,  2*i+1}$ is $1$, then it must have been produced in the previous, {\em {i.e.}},  $l$-th ,  cascade stage, and therefore, the integer represented by the binary sequence  $\basel{s}{l, \,  (2*i+1)*2^{l}+2^{l}-1} \basel{s}{l, \,  (2*i+1)*2^{l}+2^{l}-2} \,  \ldots \, \basel{s}{l, \,  (2*i+1)*2^{l}}$ can be at most $2^{2^{l}}-2$, as shown below, and hence, there is such an index $j$ as just being discussed, and the increment operation cannot further produce a carry.
\item 	The final sum is $\basel{s}{k, \,  2^{k}-1}  \basel{s}{k, \,  2^{k}-2} \, \ldots  \basel{s}{k, \,  0}$, with final carry $\basel{c}{k,\,   0}$.
\end{enumerate}

\noindent {\underline{\em{Claim}} :} The integer represented by $\basel{s}{m, \,  i*2^{m}+2^{m}-1}  \basel{s}{m,  \, i*2^{m}+2^{m}-2} \, \ldots \, \basel{s}{m, \,  i*2^{m}}$, together with carry $\basel{c}{m,\, i}$, is the result of the addition of the integers represented by $\basel{a}{i*2^{m}+2^{m}-1}  \basel{a}{i*2^{m}+2^{m}-2} \, \ldots \, \basel{a}{i*2^{m}}$ and $\basel{b}{i*2^{m}+2^{m}-1}  \basel{b}{i*2^{m}+2^{m}-2} \, \ldots \, \basel{b}{i*2^{m}}$,  for  $0 \leq i \leq 2^{(k-m)}-1$ and $1 \leq m \leq k$, as expressed by the following equation:
\begin{equation}
\basel{c}{m, \, i} * 2^{2^{m}} ~ + ~ \sum_{j = 0}^{2^{m}-1}\basel{s}{m, \,  i*2^{m}+j} * 2^{j} ~~ = ~~ \sum_{j = 0}^{2^{m}-1}\basel{a}{i*2^{m}+j} * 2^{j}  ~ + ~ \sum_{j = 0}^{2^{m}-1}\basel{b}{i*2^{m}+j} * 2^{j}
\label{Eqn-1}
\end{equation}

\noindent {\underline{\em{Proof}} :}
The claim is true for $m = 1$, by the construction in Step $3$. Now, it is assumed be true through all cascade stages up to and including $m$ and $l$, where $1 \leq m \leq l \leq k-1$. Entering the second for-loop, indexed by $0 \leq i \leq 2^{(k-l-1)}-1$, in Step $4$, it is required to show that the assertion in (\ref{Eqn-1}) holds true, for $m = l+1$. Now, by inductive hypothesis, the following is assumed to hold true, for $0 \leq i \leq 2^{(k-l-1)}-1$: 
\begin{small}
\begin{eqnarray}
&&\shiftleft \basel{c}{l, \, 2*i} * 2^{2^{l}} +  \sum_{j = 0}^{2^{l}-1}\basel{s}{l, \,  (2*i)*2^{l}+j} * 2^{j} ~~ = ~~ \sum_{j = 0}^{2^{l}-1}\basel{a}{(2*i)*2^{l}+j} * 2^{j}   +  \sum_{j = 0}^{2^{l}-1}\basel{b}{(2*i)*2^{l}+j} * 2^{j} ~~~~ \textrm{and} ~~
\label{Eqn-2}\\
&&\shiftleft \ltab \basel{c}{l, \, 2*i+1} * 2^{2^{l}}  +  \sum_{j = 0}^{2^{l}-1}\basel{s}{l, \,  (2*i+1)*2^{l}+j} * 2^{j} ~~ = ~~ \sum_{j = 0}^{2^{l}-1}\basel{a}{(2*i+1)*2^{l}+j} * 2^{j}   +  \sum_{j = 0}^{2^{l}-1}\basel{b}{(2*i+1)*2^{l}+j} * 2^{j}
\label{Eqn-3}
\end{eqnarray}
\end{small}
\lspace Multiplying both sides of (\ref{Eqn-3}) by $2^{l}$, the following is obtained:
\begin{eqnarray}
&&\ltab \ltab \basel{c}{l, \, 2*i+1} * 2^{2^{l+1}} ~ + ~ \sum_{j = 0}^{2^{l}-1}\basel{s}{l, \,  (2*i+1)*2^{l}+j} * 2^{2^{l}+j} ~~ = \nonumber \\
&&\shiftright  \sum_{j = 0}^{2^{l}-1}\basel{a}{(2*i+1)*2^{l}+j} * 2^{2^{l}+j}  ~ + ~ \sum_{j = 0}^{2^{l}-1}\basel{b}{(2*i+1)*2^{l}+j} * 2^{2^{l}+j}
\label{Eqn-4}
\end{eqnarray}
and adding the corresponding sides of (\ref{Eqn-4}) and (\ref{Eqn-2}), the following is obtained:
\begin{eqnarray}
&&\ltab \ltab \basel{c}{l, \, 2*i+1} * 2^{2^{l+1}} ~ + ~ \sum_{j = 0}^{2^{l}-1}\basel{s}{l, \,  (2*i+1)*2^{l}+j} * 2^{2^{l}+j} ~~ + ~~ \basel{c}{l, \, 2*i} * 2^{2^{l}} ~ + ~ \sum_{j = 0}^{2^{l}-1}\basel{s}{l, \,  (2*i)*2^{l}+j} * 2^{j}   \nonumber \\
&&\tab \tab = ~~~~ \sum_{j = 0}^{2^{l}-1}\basel{a}{(2*i+1)*2^{l}+j} * 2^{2^{l}+j}  ~ + ~ \sum_{j = 0}^{2^{l}-1}\basel{b}{(2*i+1)*2^{l}+j} * 2^{2^{l}+j} ~~ + \nonumber \\
&& \shiftright \sum_{j = 0}^{2^{l}-1}\basel{a}{(2*i)*2^{l}+j} * 2^{j}   +  \sum_{j = 0}^{2^{l}-1}\basel{b}{(2*i)*2^{l}+j} * 2^{j} \nonumber \\
&&\tab \tab = ~~~~ \sum_{j = 0}^{2^{l}-1}\basel{a}{(2*i)*2^{l}+2^{l}+j} * 2^{2^{l}+j}  ~ + ~ \sum_{j = 0}^{2^{l}-1}\basel{b}{(2*i)*2^{l}+2^{l}+j} * 2^{2^{l}+j} ~~ + \nonumber \\
&& \shiftright \sum_{j = 0}^{2^{l}-1}\basel{a}{(2*i)*2^{l}+j} * 2^{j}   +  \sum_{j = 0}^{2^{l}-1}\basel{b}{(2*i)*2^{l}+j} * 2^{j} \nonumber \\
&&\tab \tab = ~~~~ \sum_{j = 0}^{2^{l}-1}\basel{a}{i*2^{l+1}+2^{l}+j} * 2^{2^{l}+j}  ~ + ~ \sum_{j = 0}^{2^{l}-1}\basel{b}{i*2^{l+1}+2^{l}+j} * 2^{2^{l}+j} ~~ + \nonumber \\
&& \shiftright \sum_{j = 0}^{2^{l}-1}\basel{a}{i*2^{l+1}+j} * 2^{j}   +  \sum_{j = 0}^{2^{l}-1}\basel{b}{i*2^{l+1}+j} * 2^{j} \nonumber \\
&&\tab \tab = ~~~~ \sum_{j = 0}^{2^{l+1}-1}\basel{a}{i*2^{l+1}+j} * 2^{j}   +  \sum_{j = 0}^{2^{l+1}-1}\basel{b}{i*2^{l+1}+j} * 2^{j}
\label{Eqn-5}
\end{eqnarray}
where the last term is the result of addition of the integers represented by the pair of binary sequences $\basel{a}{i*2^{(l+1)}+2^{(l+1)}-1}  \basel{a}{i*2^{(l+1)}+2^{(l+1)}-2}\, \ldots \, \basel{a}{ i*2^{(l+1)}}$  and $\basel{b}{i*2^{(l+1)}+2^{(l+1)}-1}  \basel{b}{i*2^{(l+1)}+2^{(l+1)}-2}\, \ldots \, \basel{b}{ i*2^{(l+1)}}$ , for $0 \leq i \leq 2^{(k-l-1)}-1$. Now, either of the summands on the right hand side of (\ref{Eqn-3}) is at most $2^{(2^l)}-1$, and therefore, their sum is at most $2^{2^{l}+1}-2$, while the maximum integer that can be represented by the left hand side in (\ref{Eqn-3}) is $2^{2^{l}+1}-1$, which means that the single bit $\basel{c}{l, \,  2*i}$  can be added to the left hand side of (\ref{Eqn-3}) without an overflow, for $0 \leq i \leq 2^{(k-l-1)}-1$. Thus, by the result of the carry increment in Step 4, the following holds, for $0 \leq i \leq 2^{(k-l-1)}-1$:
\begin{small}
\begin{eqnarray}
&& \basel{c}{l+1, \, i} * 2^{2^{l}} ~ + ~ \sum_{j = 0}^{2^{l}-1}\basel{s}{l+1, \,  i*2^{l}+2^{l}+j} * 2^{j} ~~~~ =  \nonumber \\
&& \tab \tab \basel{c}{l, \, 2*i+1} * 2^{2^{l}} ~ + ~ \sum_{j = 0}^{2^{l}-1}\basel{s}{l, \,  (2*i+1)*2^{l}+j} * 2^{j}  ~ + ~\basel{c}{l, \, 2*i}
\label{Eqn-6} \\
&& \textrm{ and } \tab \sum_{j = 0}^{2^{l}-1}\basel{s}{l+1, \,  i*2^{l}+j} * 2^{j} ~~ = ~~ \sum_{j = 0}^{2^{l}-1}\basel{s}{l, \,  (2*i)*2^{l}+j} * 2^{j}
\label{Eqn-7}
 \end{eqnarray}
\end{small} 
\lspace Now multiplying both sides of (\ref{Eqn-6}) by $2^{2^l}$ and adding the corresponding sides in (\ref{Eqn-7}) to the result, and using (\ref{Eqn-5}), the following is obtained, for $0 \leq i \leq 2^{k-l-1}$:
\begin{eqnarray}
 && \ltab \ltab \basel{c}{l+1, \, i} * 2^{2^{l+1}} ~ + ~ \sum_{j = 0}^{2^{l+1}-1}\basel{s}{l+1, \,  i*2^{l+1}+j} * 2^{j}  ~~~~~ = \nonumber \\
 && \basel{c}{l+1, \, i} * 2^{2^{l+1}} ~ + ~ \sum_{j = 0}^{2^{l}-1}\basel{s}{l+1, \,  i*2^{l+1}+2^{l}+j} * 2^{2^{l}+j}  ~ + ~ \sum_{j = 0}^{2^{l}-1}\basel{s}{l+1, \,  i*2^{l+1}+j} * 2^{j} ~~~~ = \nonumber \\
 && \ltab \ltab  \basel{c}{l, \, 2*i+1} * 2^{2^{l+1}} ~ + ~ \sum_{j = 0}^{2^{l}-1}\basel{s}{l, \,  (2*i+1)*2^{l}+j} * 2^{2^{l}+j}  ~ + ~\basel{c}{l, \, 2*i} * 2^{2^{l}}  ~ + ~  \sum_{j = 0}^{2^{l}-1}\basel{s}{l, \,  (2*i)*2^{l}+j} * 2^{j}
\nonumber \\
&&\tab \tab = ~~~~ \sum_{j = 0}^{2^{l+1}-1}\basel{a}{i*2^{l+1}+j} * 2^{j}   +  \sum_{j = 0}^{2^{l+1}-1}\basel{b}{i*2^{l+1}+j} * 2^{j}
\label{Eqn-8}
\end{eqnarray}
which proves the claim for $m = l+1$. \qed
 \\
 
\noindent{\underline{\em{Circuit Complexity}} :} We estimate the number of special purpose AND-gates required for performing the carry addition operation in Step $4$.  For $1 \leq l \leq k-1$, there are $2^{(k-l)}$ many sum sequences in the input at level $l$, and, of these, only $2^{(k-l-1)}$ many, that constitute the higher precision subsequence at level $(l+1)$, are required to be incremented. Each sequence to undergo increment operation needs $(2^{l}+1)$ AND-gates. Thus the total number of special purpose AND-gates of this implementation is found as follows:
\begin{eqnarray*}
&& \sum_{l = 1}^{k-1} \left(2^{l}+1\right) * 2^{k-l-1} ~~ = ~~ \sum_{l = 1}^{k-1} \left(2^{k-1}+2^{k-l-1}\right) \\
&& \shiftright  = ~~ (k-1)*2^{k-1} +  \left(2^{k-1}-1\right) \\
&& \shiftright  = ~~ k * 2^{k-1} - 1 ~~ = ~~ \frac{N * \basel{\log}{2} N}{2}-1
\end{eqnarray*}

\section{The Usefulness of Special Purpose Circuits for Addition or Subtraction by $2^{i}$}
A processor can be furnished with a special purpose circuit for incrementing an integer represented by $N$-bit sequence by $2^{i}$, for $0 \leq i \leq N-1$.  This operation is useful in the following contexts: taking 2's complement operation, subtraction operation, increment of instruction pointer, array index computations, memory address calculation, and as a special instruction, dedicated for this purpose, similar to shift operation. The special purpose circuit is expected to take only one clock tick to perform the specified increment operation. Further, for adding an integer represented by very sparsely occupied 1-bits, the addition operation can be implemented by a sequence of such instructions. The subtraction operation by  $2^{i}$, for $ 0 \leq i \leq N-1$, can be realized complementarily. 
 
 \paragraph{\underline{Improved Parallel Adder Circuit}}
 \begin{enumerate}
\item	Let  $\basel{a}{(N-1)} \basel{a}{(N-2)} \ldots \basel{a}{0}$ and $\basel{b}{(N-1)} \basel{b}{(N-2)} \ldots \basel{b}{0}$ be the input integers in the binary form, with the convention that the most significant bit is the leftmost (and the least significant bit the rightmost).
\item In the first step, compute $N$ sums of two bits each $\basel{s}{i} = \basel{a}{i} \oplus \basel{b}{i}$ and carries
  $\basel{c}{i} = \basel{a}{i} \wedge \basel{b}{i}$, for  $0 \leq i \leq N-1$, and set $\basel{s}{N} = 0$. All the operations are performed in parallel taking only one time delay.
\item In the second step, the carries
 $\basel{c}{i}$, for $0 \leq i \leq N-1$
  are added in parallel, without conflict,
   requiring about $\frac{N*(N+1)}{2}$
    special purpose AND-gates, as follows: 
  \begin{enumerate}
   \item let, for $0 \leq i < j \leq N$,
   \begin{displaymath}
   {\mathsf {{SC}\_{AND}}}(i,\, j) ~~ =  \left \{
    \begin{array}{l}
    \overline{\basel{s}{j}} \wedge \basel{c}{i}\,, ~~~~\textrm{if} ~ j = i+1\,, ~~~~~~ \textrm{and}\\
    \overline{\basel{s}{j}}\wedge \basel{s}{j-1} \wedge \cdots \wedge \basel{s}{i+1}\wedge \basel{c}{i}\,, ~~\textrm{otherwise}    
    \end{array} \right .
    \end{displaymath}
    \item for each index $i$, where $0 \leq i \leq N-1$, if $\basel{c}{i} = 1$, there exists exactly one index $j$, such that $ i+1 \leq j \leq N$ and  $   {\mathsf {{SC}\_{AND}}} (i,\, j) = 1$, since $\basel{s}{N}$ is initialized to $0$; the uiqueness of the index $j$ can be easily deduced; moreover, $\basel{c}{l} = 0$, for $i+1 \leq l \leq j-1$, when $i+2 \leq j \leq N$, as shown below, which means that there are no more carries to be added, whose indexes are between $i+1$ and $j-1$, inclusive of both;

    \item let $j$ be the unique index as in (b) above, such that ${\mathsf {{SC}\_{AND}}}(i,\, j) = 1$ and $i+1 \leq j \leq N$;
    then ${\mathsf {{SC}\_{AND}}} (i,\, j)$ instantly complements the bit string $\basel{s}{j} \basel{s}{j-1} \ldots \basel{s}{i+1}$, for $0 \leq i \leq N-1$;
    \item the sum is $\basel{s}{N} \basel{s}{N-1} \ldots \basel{s}{0}$, with $\basel{s}{N}$ interpreted as the carry or overflow bit.
  \end{enumerate}
  
\end{enumerate}

\noindent{\underline{\em{Proof of Correctness of the Algorithm}}:}  Let $0 \leq \basel{i}{1} < \cdots < \basel{i}{r} \leq N-1$ be the distinct indexes, such that $\basel{c}{\basel{i}{l}} = 1$, for  $1 \leq l \leq r$, for some positive integer $r$,  and $\basel{c}{i} = 0$, for $i \not\in \{\basel{i}{1}, \ldots, \basel{i}{r} \}$, where $0 \leq i \leq N-1$ and $1 \leq r \leq N$. If $r =1 $, then $\basel{c}{\basel{i}{1}}$ is the only carry to be added, and this case is easily handled by the algorithm. Let   $2 \leq r \leq N$. The main point in the proof is that the addition operation of a carry $\basel{c}{\basel{i}{l}}$  does not affect the addition operation of the carry $\basel{c}{\basel{i}{l+1}}$, for $0 \leq l \leq r-1$, as observed in the following.  The bit $\basel{s}{\basel{i}{l+1}}$ must be $0$, because $\basel{c}{\basel{i}{l+1}} = 1$ and $\basel{c}{\basel{i}{l+1}}\basel{s}{\basel{i}{l+1}}$,  being the result of adding only two bits, $\basel{a}{\basel{i}{l+1}}$ and $\basel{b}{\basel{i}{l+1}}$, cannot be the bit string $11$,  for $1 \leq l \leq r-1$. Thus, there exists an index $\basel{j}{l}$, such that $\basel{i}{l}+1 \leq \basel{j}{l} \leq \basel{i}{l+1}$ and ${\mathsf {{SC}\_{AND}}} (\basel{i}{l},\, \basel{j}{l}) = 1$, for $1 \leq l \leq r-1$. Now, since there are no carries of $1$s, whose indexes between $\basel{i}{l}+1$ and $\basel{i}{l+1}-1$, inclusive of both, when $\basel{i}{l} + 2 \leq \basel{i}{l+1}$, the complementation of the string $\basel{s}{\basel{j}{l}}\basel{s}{\basel{j}{l}-1} \cdots \basel{s}{\basel{i}{l}+1}$ is equivalent to adding $1$ to the corresponding integer represented by it, without affecting the carry addition of  $\basel{c}{\basel{i}{l+1}}$, for $1 \leq l \leq r-1$. The last carry $\basel{c}{\basel{i}{r}}$ is added, as if it were lone carry to be added. \qed
\\

It may be observed that addition of two $(2N)$-bit integers takes only 3 time delays by means of two $N$-bit adders as just described. Two lower and higher significant $N$-bit integers are added, and if a carry is produced by the addition operation of the two lower significant $N$-bit integers, then it is added to the sum of the two higher significant $N$-bit integers, in just one time delay. The last step may require additional $N$ special purpose AND-gates, for the addition operation by $1$, when the initialization at the leaf node is two bits at a time, by means of associative memory units. Thus, the total number of special purpose AND-gates could be about $2 \times \frac{N(N+1)}{4}+N = \frac{N(N+3)}{2}$, for addition of two $(2N)$-bit integers, in three time delays. The application for multiplication of two $N$-bit integers is discussed in the next section.

In the first attempt algorithm described in the previous section, we started at leaf node with sums of two bits of $a$'s and $b$'s each, at a time. If we assume a similar initialization to compute the sum $s$ and carry $c$ bits, we could reduce the space required by a factor of $2$ for the special purpose AND-gates, in the algorithm just described in this section.  Another possibility for reduction of the number of special purpose AND-gates, for the sake of economy, is to consider a two-level cascaded implementation. In the first cascade stage, about $\sqrt{N}$ blocks are taken for addition in parallel, each block consisting of again about $\sqrt{N}$ sum and carry bits. In this circuit design, the number of special purpose AND-gates in the first cascade stage would be about $\sqrt{N} * \frac{\sqrt{N}(\sqrt{N}+1)}{2} = \frac{N(\sqrt{N}+1)}{2}$. In the second cascade stage, there are about $\sqrt{N}$ carry bits to be added, which would require about $\frac{\sqrt{N}*(N-\sqrt{N}+2)}{2}$ special purpose AND-gates, because the least significant $\sqrt{N}$ bits do not affect the carry addition to the higher significant $(N-\sqrt{N}+1)$ bits. Thus, the total number of special purpose AND-gates could be about $(N˘+1)\sqrt{N}$. Combined with the previous observation, {\em {i.e.}}, starting with two bits of $a$'s and $b$'s to get the  $s$  and  $c$ bits in the initialization step, it is possible to realize a $(2N)$-bit integer adder performing the addition operation  in three clock ticks, requiring about  $(2N+1)\sqrt{N}-\frac{N}{2}$ special purpose AND-gates. The estimates are as follows: 
\begin{enumerate}
\item there are $N$ carries to be added after the initialization step; 
\item in the first cascade stage, there are $2\sqrt{N}$ sum bits and $\sqrt{N}$ carries, in each block,  which would need 
 $\sqrt{N} * \frac{2\sqrt{N}(2\sqrt{N}+1)}{4} = \frac{N(2\sqrt{N}+1)}{2} = N\sqrt{N}+\frac{N}{2}$ special purpose AND-gates; and 
 \item in the second cascade stage, there are $\sqrt{N}$ carries, needing an average of $\frac{(2N-2\sqrt{N}+2)}{2} = N-\sqrt{N}+1$ special purpose AND-gates per one carry bit, resulting in an estimate of $N\sqrt{N}-N+\sqrt{N}$ special purpose AND-gates.
\end{enumerate} 
  Thus, the total number of special purpose AND-gates in this construction would be about $2N\sqrt{N}-\frac{N}{2}+\sqrt{N} = (2N+1)\sqrt{N}-\frac{N}{2}$ carries.  For typical numbers, if $N = 64$, then  $(2N+1)\sqrt{N}-\frac{N}{2} = 1000$, as compared to $(N(N+3))/2 = 2144$, required by the circuit without the space reduction by  two-stage cascaded implementation, both circuits taking only three clock ticks to add two $(2N)$-bit --  {{\em{i.e.}}, two $128$-bit -- integers. On a $64$-bit processor, $128$-bit integer adder is needed for multiplication operation. The first attempt design circuit of the previous section would need $(128*7)/2-1 = 447$ special purpose AND-gates, performing the addition of two 128-bit integers in about 7 clock ticks, while a two-stage cascade circuit would need about $1000$ special purpose AND-gates, to repeat, performing the addition of two 128-bit integers in $3$ clock ticks. In Slide 83 of \cite{Rudich:2004}, it is stated that the Pentium processor performs the 32-bit integer addition in $11$ gate delays.

\section{Multiplication of Two Integers in Binary Representation}
The time delay of multiplication of two $N$-bit integers is determined mostly by the time delay of addition of $(2N)$-bit integers, requiring at least one $(2N)$-bit adder and consolidation circuits that reduce a larger number of integers to a smaller number of integers for addition, such that the sum of the integers, before and after consolidation, is the same. For each index $i$, a Cauchy sum of product is formed, which corresponds to the coefficient of $2^{i}$, for $0 \leq i \leq 2N-1$. Since there are at most $N$ products of two bits in each sum, they are added in $\basel{\log}{2} N$ stages, to get $2N$ coefficients represented by at most  $\basel{\log}{2} N$ bits each. Then, the bit-planes of the coefficients are rearranged, similar to rearranging the order of summation of a doubly indexed sum, into  $\basel{\log}{2} N$ integers of at most $2N$ bits, with $(N+1)$ quantization levels, which can be classified by $(N+1)$ comparators (Chapter 7 of \cite{MT:1965}). The quantization intervals are recognized by two adjacent voltage levels. The voltages of the bits in a column corresponding to the same place of a nonnegative integer power of $2$ are connected in series, to get the sum of voltages, which encodes the number of $1$s in the column. If the bits are sensitive to current measurements, then they are added in parallel, to form the sum of currents. The common junction point is connected to the ground by an additional resistor. Thus, in any case, the sum of the voltages is measured at a particular junction point. The sum falls (after accounting for small errors and fluctuations) somewhere in the middle of exactly one quantization interval, which is recognized by the conjugation of the conditions that ($i$) the upper limit voltage is larger,  and ($ii$) the lower limit voltage is smaller than the sum of the voltages in a column. The conjunction of the two conditions is fed to a switching circuit (Chapter 8 of \cite{MT:1965}), which switches an associative memory entry containing the bit pattern that encodes the integer to count the number of $1$s ub the column. Thus, the sum of $\nu \geq 3$ integers can be reduced to a sum of  $\lfloor \basel{\log}{2} \nu \rfloor +1$ integers, in a constant number of (which may be two) clock ticks. However, when the number of integers to be added falls to a small number (such as below 6), the consolidation method described in Slide 45 of \cite{Rudich:2004} may be faster than the quantizer circuit. The quantizer based consolidation method achieves higher speed, when the number of integers to be consolidated is larger than a prescribed number, and as such may be qualified to be called optimal, owing to its constant time operational performance. The final two integers after the consolidation stages are added to get the integer which is the product of the two integers, given as input in the beginning. 

The consolidation operation is illustrated for the $64$-bit multiplication. Initially, there are $64$ integers to be added, which are aligned properly adjusting for the respective binary places. Two cases are described for comparison: one with only $3$-bit to $2$-bit consolidation circuits described in Slide 45 of \cite{Rudich:2004}, and the other with quantizers for about two stages followed by $3$-bit to $2$-bit consolidation circuits described in Slide 45 of \cite{Rudich:2004} in the remaining stages, until both reduce the sum of the initially given $64$ integers into a sum of $2$ integers, where the latter could be $128$-bit long, unlike in the input, which are at most $64$-bit long. The quantizer is assumed to take two clock ticks to produce the required integers, as follows: in the first clock tick, the lower and upper bounds of interval of quantization are detected, consequently initiating the corresponding switching circuit, and in the second clock tick, the initiated switching circuit activates an associative memory unit, which places the contents in appropriate places, taking care also of the binary places, positioning the resulting integers as in a staircase, for the next stage. The circuit initialization phase is sensitive to the leading or trailing edge of a switching (initiating) pulse, giving the pipeline or cascade effect, which is partly folded into (overlapped with) the duration of the switching pulse. The edges are not always sharp or crisp, and edge sensitivity is exploited for gaining speedup in cascading (during both feed-forward and feedback stages of) compound circuits. The measurements for settling time for the overall circuit are explicitly performed, by trying out its response for various pulses that arise in typical (empirical) situations.

\begin{description}
\item[{\underline{(A) \bf With only $3$-bit to $2$-bit consolidation.}}]  The numbers of integers to be consolidated in a sequence of stages taking only one clock tick per stage are as follows (where the serial number stands for the clock tick offset number): (1) $64$ to $43$ (with only $63$ to $42$ consolidation and one integer left out), (2) $43$ to $29$ (with only $42$ to $28$ consolidation and one integer left out), (3) $29$ to $20$ (with only $27$ to $18$ consolidation and two integers left out), (4) $20$ to $14$ (with only $18$ to $12$ consolidation and two integers left out), (5) $14$ to $10$ (with only $12$ to $8$ consolidation and two integers left out), (6) $10$ to $7$ (with only $9$ to $6$ consolidation and one integer left out), (7) $7$ to $5$ (with only $6$ to $4$ consolidation and one integer left out), (8) $5$ to $4$ (with only $3$ to $2$ consolidation and two integers left out), (9) $4$ to $3$ (with only $3$ to $2$ consolidation and one integer left out) and (10) $3$ to $2$ consolidation, taking $10$ clock ticks to complete the task. The overall consolidation factor for consolidating $64$ integers into $2$ integers is $32$, and with a consolidation factor of $\frac{3}{2}$ per stage, the lower bound for the number of stages is $\lceil \basel{\log}{{3}/{2}} (32) \rceil  = \lceil 8.547... \rceil = 9$. The overrun of the number of stages is caused by the nondivisibility of the number of integers to be consolidated by the integer $3$ in some stages.

It may be observed that,  with required quantizers to add up $14$ bits to produce $4$-bit integers in binary representation,  steps (5) through (8) can be replaced with a single quantizer step, which may take two clock ticks to perform this particular subtask, saving two clock ticks. As another opportunity, again with required quantizers to add up $7$ bits to produce $3$-bit integers in binary representation, for instance, steps (7) through (9) can be replaced with a single quantizer step, which may take two clock ticks to perform this particular subtask, but saving just one clock tick. 

\item[{\underline{(B) \bf With quantizers and $3$-bit to $2$-bit consolidation.}}]
 The numbers of integers to be consolidated in a sequence of stages taking one or two clock ticks per stage, depending on the particular stage, are as follows (the serial number marking for the end of the clock tick offset number): (2) $64$ to $7$ (with $63$-bit to $6$-bit consolidation based on quantizers, taking two clock ticks, and one integer left out), (4) $7$ to $3$ (with $7$-bit to $3$-bit consolidation based on quantizers, taking two clock ticks),  and (5) $3$ to $2$ consolidation (with only $3$-bit to $2$-bit consolidation, taking one clock tick), taking $5$ clock ticks to complete the task. 
\end{description}

For  the overall time needed,  $5$ clock ticks for consolidation of $64$ to $2$ integers of at most $128$ bits each, added to about $3$ clock ticks for the addition of the two $128$-bit integers, to get the final result of multiplication of the two input $64$-bit integers in about $8$ clock ticks, in case {\bf (B)}, and  about $9$ clock ticks obtained by the theoretical lower bound for consolidation of $64$ to $2$ integers of at most $128$ bits each, added to about $15$ clock ticks for the addition of the two $128$-bit integers,  to get the final result of multiplication of the two input $64$-bit integers in about $24$ clock ticks, in case {\bf (A)}. Thus, the speedup factor is at least $\frac{24}{8} = 3$.

In the following discussion, the circuit complexity for the two cases discussed above is estimated. The initial $64$ number of $64$-bit integers are arranged in a parallelogram staircase, in the standard presentation. They can be arranged to foom a nabla ($\nabla$) or Delta ($\Delta$) shape staring at $127$-bit integer in the first row, followed by $125$-bit integer in the second row and so on, until $1$-bit integer in the last ($64$-th) row. In the first stage, since $64$ itself is not divisible by $3$, there are $63$ rows to be consolidated, and $121$ number of $3$-bit to $2$-bit consolidation circuits, required in the second row, followed by $115$ number of $3$-bit to $2$-bit consolidation circuits, required in the fifth row, until one $3$-bit to $2$-bit consolidation circuit, in the $62$-nd row, skipping two rows in between, with $6$ circuits less in succession. These consolidation circuits must perform in parallel in the first stage at least. This number can also be arrived at by observing that $21$ rows of $3$-bit to $2$-bit consolidation circuits are required to consolidate $63$ rows to $42$ rows in the first step.  Thus, there are $\sum_{i = 0}^{20} (6*i+1) = 1+ 7 + \cdots + 121 = 21 * 61 = 1281$ number of $3$-bit to $2$-bit circuits (associative memory units) required, in case {\bf (A)}, each circuit containing $8$ entries of $2$-bit associative memory. Now, in case {\bf (B)}, in addition to $128$ number of $3$-bit to $2$-bit consolidation circuits in the final consolidation stage, the number of $63$-bit to $6$-bit quantizers needed is about $128$, with possible reuse in the second stage, and if no reuse is possible, another $128$ number of $7$-bit to $3$-bit quantizers in the second consolidation stage are needed. For comparison, $128$ number of $63$-bit to $6$-bit quantizers hold $64*128 = 8192$ associative memory entries of $6$-bits each, while $1281-128 = 1153$ number of $3$-bit to $2$-bit consolidation circuits hold  $1153 * 8 = 9224$ number of $2$-bit associative memory entries. If reuse of the quantizers in the second stage is possible, the associative memory space requirement in case {\bf (B)} is less than $3$ times that in case {\bf (A)}, with a speedup factor of at least $3$. It may be observed that the well-known Amdahl's law for speedup bound is applicable for the same programs or circuits, when executed in parallel by replication of resources. An interesting situation is when different tasks together require some resources in total, which can be allocated to them to execute in parallel, without requiring any additional resources. Quantizers are more commonly well-known in the analog-to-digital (ADC) converters. However, the inputs to the quantizers in this section take only finitely many discrete values, and the required precision for the lower and upper bounds of the interval of quantization for the sum offers considerable tolerance for accounting for small errors and fluctuations in the current or voltage measurements taken at the input.

\end{document}